\documentstyle[mnras_citedal,psfig]{mn2e}

\newif\ifAMStwofonts

\def\f0{$f_0({\bf v})$}
\def\Omegap{{\Omega_{\rm p}}}

\def\ROLR{R_{\rm OLR}}
\def\fsing{f_{\rm s}}
\def\fexp{f_{\rm exp}}

\def\Rc{R_{\rm c}}
\def\Lc{L_{\rm c}}
\def\Psibar{\Psi_{\rm bar}}
\def\Psispiral{\Psi_{\rm spiral}}

\def\Psip{\Psi_{\rm p}}
\def\xr{x_{\rm r}}
\def\yr{y_{\rm r}}
\def\fx{f_x}
\def\fy{f_y}
\def\ds{\displaystyle}
\def\kms{\ km s$^{-1}$}

\def\sigmau{{$\sigma_{\rm u}$}}
\def\sigmav{$\sigma_{\rm v}$}
\def\sigmauv{$\sigma_{\rm uv}$}
\def\<#1>{\langle#1\rangle}
\def\pcite#1{[ref]}

\title[Patterns in The Outer Parts of Galactic Disks]
       {Patterns in the Outer Parts of Galactic Disks}

\author[Chakrabarty D.]
       {D. Chakrabarty,\\
        Dept. of Physics and Astronomy, Rutegrs University\\
        136, Frelinghuysen Road, Piscataway, NJ 08854-8019\\
}
\date{\today}

\begin{document}

\maketitle

\label{firstpage}

\begin{abstract}
This paper describes test particle simulations of the response of the outer
parts of Galactic disks to barring and spiral structure.  Simulations
are conducted for cold Mestel disks and warm quasi-exponential disks
with completely flat rotation curves, subjected to pure quadrupoles
and logarithmic spirals. Even though the starting velocity
distributions are smooth, the end-points of the bar simulations show
bimodality and multi-peaked structures at locations near the outer
Lindblad resonance (OLR), although spirality can make this smoother.
The growth of a bar may cause the disk isophotes to become boxy at the
OLR, as stars accummulate particularly along the minor axis. The
growth of a bar is also accompanied by substantial heating of the disk
stars near the OLR. For the growth of a $10^{10} M_{\rm \odot}$ bar,
the radial velocity dispersion is typically quadrupled for initially
cold disks (initial \sigmau $\sim 10$ \kms), and typically doubled for disks
with final \sigmau $\sim 45$ \kms. Simulations performed of the growth and
dissolution of bars give very similar results, demonstrating that
the heat once given to disk stars is very difficult to remove.

\end{abstract}


\section{Introduction}
\label{sec:intro}

\noindent
The traditional approach to stellar dynamics, which stems from the
work of Eddington and Jeans in the early years of the twentieth
century, assumes that the phase space distribution of stars is smooth.
From this innocent enough assumption, the remarkable and powerful
Jeans (1915) theorem follows. This constrains the phase space
distribution function to depend on the isolating integrals of motion
only.

Nowadays, it is less clear that the assumption of a smooth
distribution which underpins classical stellar dynamics is at all
justified.  For example, it has been realised that the local velocity
distribution in the Galactic disk is deformed by stellar streams and
moving groups (e.g. Eggen 1996a, Eggen 1996b). Members of such
kinematically cold streams move with similar space
motions. \scite{agris} presented convincing evidence that the local
distribution of stars is strongly bimodal. He identified two streams
of stars, the Hyades and the Sirius superclusters.  More recent
pictures of the local stellar velocity distribution, obtained by
\scite{walterdf} with the {\it Hipparcos} data, revealed extensive
clumps of structure in the $(U,V$) plane (i.e., the horizontal
velocity plane) of disk stars.

The aim of this paper is to investigate the phase space structures
that form in galactic disks using test particle simulations. Such
structure may be generated by a number of dynamical mechanisms. As
\scite{agris} originally suggested, the growth of a bar may cause the
trapping of stars in the vicinity of the outer Lindblad resonance and
this is a possible explanation of the structures observed in the solar
neighbourhood. Both the growth and the dissolution of bars are
processes that are likely to have occurred in many disk galaxies
(e.g. \cite{wynjim}, \cite{liaiau}) and it is interesting to look for
possible diagnostics. Second, any spirality is also likely to buffet
the stars and generate characteristic structures in phase space.

The paper is organised as follows. In Section 2, some details of the
numerical procedure are given. The calculations are all test-particle
simulations, as is applicable to the study of the outer parts of disk
galaxies where the potential of the dark halo is important. Sections
3, 4, 5 and 6 study the effects of a bar and spirality, (alone as well
as acting in concert with each other), on physical quantities such as
the velocity distribution, dispersion, disk heating and surface
density.  Section 7 considers the growth and dissolution of bars and
spiral arms.

\section{Simulations}

\label{sec:diskpot}

\subsection{Equilibrium Models}
\label{sec:equimodel}

\noindent
I assume that the equilibrium disk is scale-free with a flat rotation
curve. The form of the potential is:
\begin{equation} 
\Psi = -v_{0}^{2}\ln\left(\frac{R}{R_{0}}\right).
\end{equation}
Here, $v_0$ is the constant amplitude of the circular velocity curve,
while $R_0$ is a characteristic lengthscale. The circling frequency
$\Omega$ and the epicyclic frequency $\kappa$ are
\begin{equation}
\Omega = {v_0 \over R}, \qquad \kappa = {2v_0 \over R}.
\end{equation}
The outer Lindblad resonance (OLR) occurs at
\begin{equation}
\ROLR =  \left(1 + {1 \over \sqrt{2}}\right){v_0\over \Omegap}.
\end{equation} 
I always scale the pattern speed of the non-axisymmetric perturbation
($\Omegap$) to unity, so that corotation (CR) occurs at a
Galactocentric radius of $R_0$. Without loss of generality, both $R_0$ 
and $v_0$ are often set to unity.

The self-consistent distribution functions (hereafter DFs) depend on
the isolating integrals of motion only. These are the binding energy
per unit mass $E$ and the angular momentum component perpendicular to
the plane of the disk $L_z$.  The self-consistent DFs are (e.g.
Chapter~4 in \cite{bible}, \cite{wynpower})
\begin{equation}
\fsing(E,L_z) = \tilde{C} L_z^\gamma \exp \left( -(\gamma+1){E/v_0^2} \right).
\label{eqn:fsingMestel}
\end{equation}
Here, $\tilde{C}$ is a normalisation constant, whose value can be
found in \scite{wynpower}.  The velocity anisotropy constant $\gamma$
prescribes the radial and tangential second moments \sigmau$^2$ and
\sigmav$^2$ of the stars. They are related to the anisotropy parameter
$\gamma$ by:
\begin{equation}
\sigma_{\rm u}^2 = {\ds
                 {{v_0^2} \over {1+\gamma} } }, \qquad\qquad
\sigma_{\rm v}^2 = {\ds {v_0^2}}.
\label{eqn:sigmaexpo}
\end{equation}
The exponential surface density profile of a disk galaxy like ours is
well-reproduced by the ``doubly cut-out power law distribution
functions'' (\cite{jerrywyn}). The equilibrium distribution function
of such a doubly cut-out disk is given as:
\begin{equation}
\fexp(E,L_z) = H(L_z) \fsing(E,L_z).
\label{eqn:fexp}
\end{equation}
Here, $\fsing$ represents the self-consistent DF given in
(\ref{eqn:fsingMestel}), while $H(L_z)$ is the
cut-out function:
\begin{equation}
H(L_z) = \frac{L_z^N \Lc^M}{[L_z^N+(v_{\beta}R_0)^N][L_z^M+L_c^M]}.
\end{equation}
Here, $N$ and $M$ are the inner and outer cut-out indices. They are
chosen as positive integers and they control the sharpness of the
inner and outer cut-outs respectively.  The constant $\Lc$ is the
angular momentum of a circular orbit at the cut-out radius $\Rc$.
Here, $v_{\beta}$ is related to the rotation curve via:
\begin{equation}
v_{c}^{2}(R) = v_{\beta}^{2}\left(\frac{R_{0}}{R}\right)^{\beta}.
\end{equation}
and $\beta$ is the index that controls the fall-off of the circular
velocity with galactocentric radius.

Computation of the surface density corresponding to the doubly cut-out
DF involves numerical integration. By choosing the relevant parameters
carefully, the surface density profile of the disk can be made to look
exponential. Such models are referred to as ``quasi-exponential''.

\begin{figure}
\centerline{
\psfig{figure=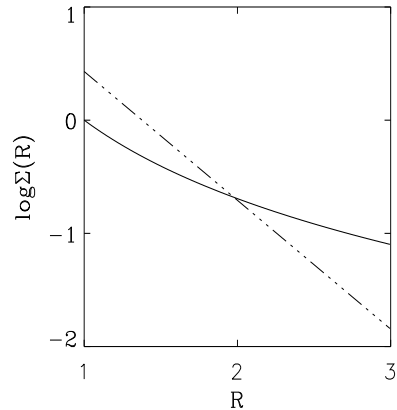,height=6.25truecm,width=6.2truecm}}
\caption{The logarithm of the surface density is plotted as a function of 
radius. The quasi-exponential disk is characterised by a surface density
profile that is well approximated by an exponential; it is represented
by the dash-dotted straight-line. The $1/R$ surface density
profile of the Mestel disk is shown by the solid line.}
\label{fig:logsurf}
\end{figure}

In this paper, two models are used as standards.  The first is a cold
Mestel disk ($\gamma =600$, $\beta =0$), whose DF is given by
eq.~\ref{eqn:fsingMestel}. This has a flat rotation curve.  The second
is a warm quasi-exponential disk ($\gamma = 45$, $\beta =0$, $M = 3$,
$N = 1$, $\Rc =2$), whose DF is given by eq.~(\ref{eqn:fexp}). Here,
the rotation curve is still flat, as the halo provides the missing
forces.  The surface density of the models is shown in
Figure~\ref{fig:logsurf}.

\subsection{Perturbations}
\label{sec:perturbation}

\noindent
The general procedure is to set up initial positions and velocities
corresponding to a smooth, equilibrium model. This equilibrium is
subjected to perturbing forces $\Psip (\xr, \yr)$ caused by a rotating
bar or spiral pattern.  The Cartesian coordinates in the rotating
frame ($\xr, \yr$) of the perturber, (with a pattern speed of
$\Omega_{\rm p}$) are related to those in the inertial frame ($x,y$)
by a rotational transformation about the angle $\Omegap t$. The
perturbation potential is steady in time in the frame rotating with
pattern speed $\Omegap$, so the components of the disturbing forces
referred to the inertial frame ($\fx, \fy$) are also related to the 
components in the rotating frame via a similar transformation.

This gives us the equations of motion as
\begin{eqnarray}
{\ddot x} &= -{\displaystyle x v_\beta^2 R_0^\beta \over \displaystyle
R^{\beta+2}} + \fx,
\nonumber \\
{\ddot y} &= -{\displaystyle y v_\beta^2 R_0^\beta \over \displaystyle
R^{\beta+2}} + \fy.
\end{eqnarray}
The orbits are then computed with either standard Runge-Kutta or leapfrog
integrators.

After the perturbation reached full strength and the stellar system
settled down, positions and velocities of the stars in the sample were
recorded. I exploit the time averages theorem (\cite{bible}), which
says that time averages are equivalent to phase averages for a
steady-state system. Typically, $10^3$ orbits were evolved under the
perturbing forces. These were each sampled randomly $10^3$ times after
the potential had settled to its steady-state endpoint.

In order to compute the velocity moments and distributions, the
spatial coordinates of the stars are placed on a grid that is regular
in radius and azimuth. The kinematical data in each cell are used to
construct the local bivariate velocity frequency distribution. Results
are often shown as ($U$, $V$) contour plots, where $U$ is the radial
velocity (measured positive towards the Galactic Center) with respect
to the Sun, whilst $V$ is the azimuthal velocity (measured positive in
the direction of Galactic rotation) with respect to the Sun. The solar
peculiar motion is $U_\odot = 5$\kms and $V_\odot = 10$ \kms (see
e.g. \cite{bm}).
\begin{figure*}
\centerline{
\psfig{figure=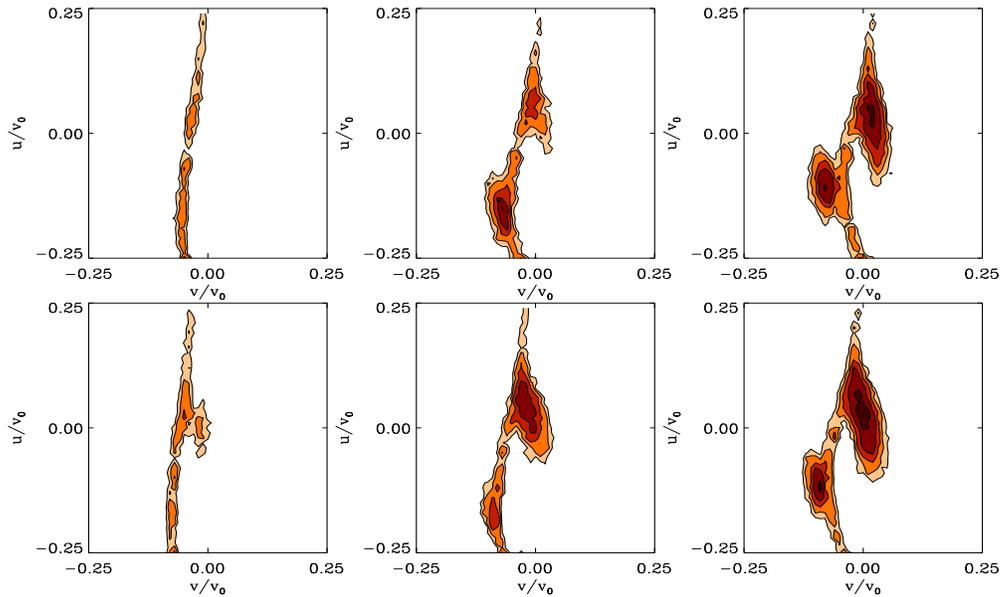,height=8.5truecm,width=13.2truecm}}
\caption{Plots of the velocity distributions in the ($U,V$) plane
after the growth of a bar. The initial model is a cold Mestel disk.
It is perturbed by a growing bar, the maximum strength of which is
1.8$\%$ of the gravitational field of the background disk, at the
OLR. The upper three panels show the ($U,V$) distributions just within
OLR at azimuths of roughly $25^\circ, 45^\circ$ and $65^\circ$. The
lower three panels show the distributions just outside the OLR and at
the same azimuths. The velocity distribution function decreases
outwards as represented by the gradient in the shading.}
\label{fig:standardmodel}
\end{figure*}
%
\begin{figure*}
\centerline{
\psfig{figure=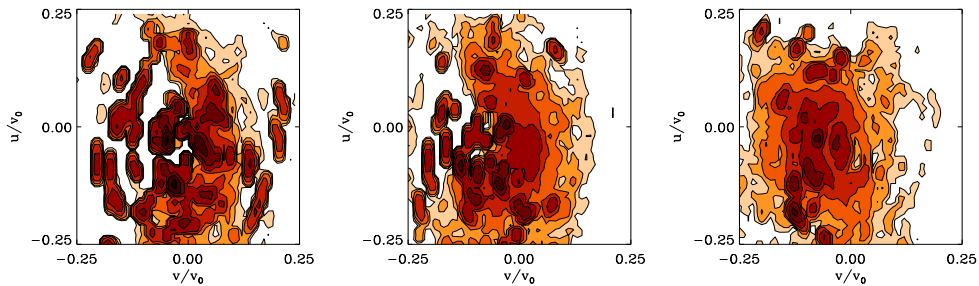,height=4.125truecm,width=13.2truecm}}
\caption{ Plots of the velocity distributions in the ($U,V$) plane after
a bar has perturbed an initially warm, quasi-exponential disk.
The maximum bar field strength is 2.4$\%$ of the background field, at
the OLR. The left panel shows the distribution at the OLR location
while the other two panels are at radii further out. The observer is
at an azimuth of 25$^\circ$ in all these diagrams. The distribution
diagrams have not been smoothed here. All other parameters are the
same as Figure~\ref{fig:standardmodel}. }
\label{fig:exponentialmodel}
\end{figure*}
\subsubsection{Bar}

The first example presented below concerns the growth and dissolution
of bars.  As a simple model of the gravity field of a bar, I use a
rigidly rotating quadrupole:
\begin{equation}
\Psibar = \epsilon\frac{\cos(2\phi)}{R^3}
\label{eqn:barpot}
\end{equation}
Here, $\phi$ is the azimuthal angle and $\epsilon$ is a measure of the
strength of the bar. $\zeta$ is used to denote the ratio of the maximum
of the radial force of the perturbation to that of the equilibrium
potential at the OLR.  The bar is grown adiabatically using the rule
\begin{equation}
\epsilon  =\cases{\epsilon_0 \tanh^4 \Bigl( {\ds t\over \ds \tau} \Bigr)&
            if $t \ge 0$,\cr
            0&otherwise.\cr}
\label{eqn:turnon}
\end{equation}
Typically, orbits are computed in the axisymmetric potential for a
time $2\tau$, then the bar is turned on and grows to its maximum
strength over a growth time $4\tau$.  After this, orbits are recorded
over a period of $2\tau$. To ensure adiabaticity, $\tau$ is chosen
such that the growth time is much larger than the revolution time of the
bar.

\subsubsection{Spiral Pattern}
\label{sec:spiralchoice}

The second example of a perturbation is weak spirality, modelled as a
logarithmic spiral. I use a spiral pattern with a pitch angle
$i=15^{\circ}$ as motivated by observations of the Milky Way
(\cite{vallee95}).  Typically a 4-armed spiral pattern is used, with a
pattern speed that is 21/55 times that of the bar, which is within the
range thought to be valid for the Milky Way (\cite{amaral}).  This
ratio places the ILR of the 4-armed spiral at the physical location of
the OLR of the bar.  On occasions, I also use 25/55 as the ratio of
pattern speeds, which places the ILR of the spiral $\sim 20 \%$ inward
of the OLR of the bar.

The perturbation potential has the following form
(eg. \cite{wynjenny}):
\begin{equation}
\Psispiral = \epsilon  K(\alpha,m)R_{0}e^{i(m\phi-\Omega_{p}t)}
\left(\frac{R}{R_{0}}\right)^{i\alpha-\frac{1}{2}}.
\label{eqn:spiralpot}
\end{equation}
Here $K(\alpha, m)$ is the Kalnajs Gravity function defined as:
\begin{equation}
K(\alpha, m) = \ds{\frac{1}{2}
 \frac{\Gamma\left(\frac{1}{4}+\frac{m}{2}+\frac{i\alpha}{2}\right)
       \Gamma\left(\frac{1}{4}+\frac{m}{2}-\frac{i\alpha}{2}\right)}
      {\Gamma\left(\frac{3}{4}+\frac{m}{2}+\frac{i\alpha}{2}\right)
       \Gamma\left(\frac{3}{4}+\frac{m}{2}-\frac{i\alpha}{2}\right)}
                  }
\end{equation}
In Equation~\ref{eqn:spiralpot}, $\epsilon$ is again a measure of the
strength of the spirality. Again, this is related to the quantity
$\zeta$, which we define to denote the ratio of the
maximum of the radial force of the perturbation to that of the
equilibrium potential at the ILR. For the same value of $\epsilon$,
the maximum radial force due to the bar at its OLR is $\sim 25 \%$
greater than that due to the spiral patten at its ILR.  The spiral
pattern is also grown adiabatically using the rule~(\ref{eqn:turnon}).

\subsubsection{Bar and Spiral Pattern}

As the third example of a non-axisymmetric perturbation, I have used
both the quadrupolar bar and the standard logarithmic spiral pattern to
perturb a background disk.  At the OLR due to the bar, the value of
$\zeta$ produced by the bar and spiral acting together, is about $1.6$
times that due to the bar alone and about $1.28$ times due to the
spiral alone.

The orbital parameters are recorded in the rotating frame of the bar,
in which the spiral pattern is not stationary. This adds an extra
element of complexity to these simulations.  In these smulations, if
orbital data is recorded at random time points, (as is done in the
bar-only or spiral-only simulations), then the effects of the
perturbations are completely washed out, as the bar and spiral pattern
are not synchronised. Thus, I record data only at those times when
the bar and spiral pattern are at the same relative
configurations. This is done by checking the phase of the potential of
the bar against that of the spiral, on the corotation circle.

\section{Velocity Distributions}
\label{sec:baruv}

\noindent
Figs.~\ref{fig:standardmodel} - \ref{fig:exponentialmodel} show the
velocity distributions at the end of simulations for a cold Mestel
disk and a warm quasi-exponential disk respectively. These disks have
been perturbed by a bar; when added to the Mestel disk, the
strength parameter for the perturbation is $\zeta = 0.018$ and when
imposed on the quasi-exponential disk, it is $0.024$. These numbers
are chosen to bring out illustrative features in the results.  In
Figure~\ref{fig:standardmodel}, the top panels show a location just
within the OLR, the bottom panels a location just outside the OLR. The
velocity structure is shown at three angles to the bar's major axis,
namely $25^\circ$, $45^\circ$ and $65^\circ$.  In
Fig.~\ref{fig:exponentialmodel}, I only show plots at different radii.

Broadly speaking, it is observed that for the cold disk, the most
distinctive feature of the corresponding velocity distribution is the
presence of two clumps at most observer locations.  As \scite{agris}
originally pointed out, and \scite{walterolr} adumbrated, these clumps
are built up from the families of stars on orbits that are aligned and
anti-aligned with the long axis of the perturber. In the warm disk,
the different stellar families are much less sharply defined on the
$(U,V)$ plane. A larger portion of the velocity space is covered in the
warm disk simulations.
\begin{figure}
\centerline{
\psfig{figure=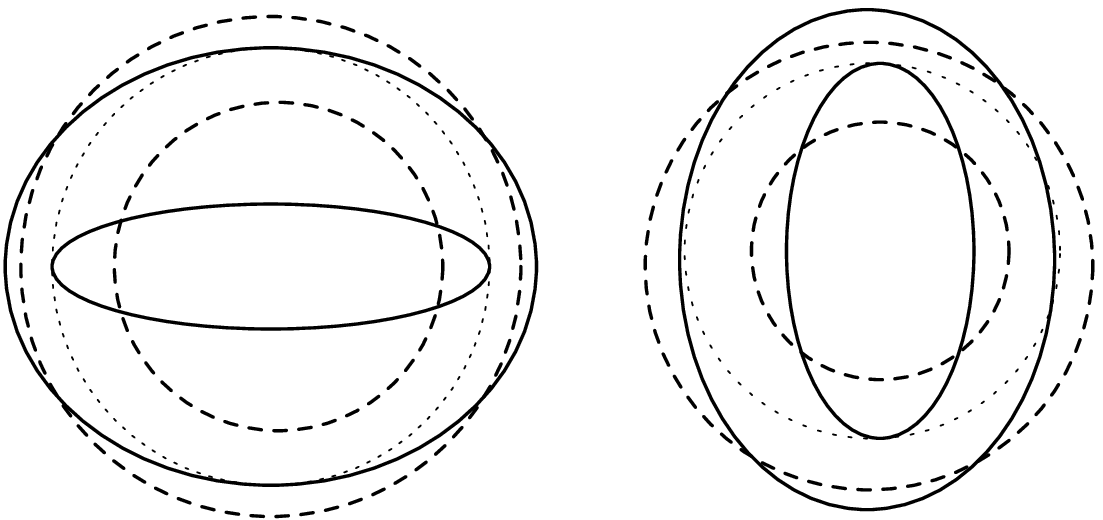,height=4.0truecm,width=7.5truecm}}
\caption{At the same radius, (shown in dotted lines) the anti-aligned
orbits (figure in solid lines, on the right) visible at low azimuths
are sired by larger unperturbed orbits (shown by dashed lines) than
those at high azimuths. Also, aligned orbits (figure in solid lines on
the left) visible at low azimuths have relatively smaller
progenitors. Azimuth is considered to increase with movement away from
the bar major axis, in the first quadrant.}
\label{fig:samefam}
\end{figure}

In terms of location in the $(U,V)$ plane, the clump representing
stars on aligned orbits is centered around $U \approx 0$ and these
stars generally have positive $V$ values. The clump representing the
anti-aligned orbits generally lies in the quadrant $U < 0$ and $V <
0$. The distinction between these clumps becomes blurred as the
background disk gets hotter. Roughly speaking, since the progenitors
of anti-aligned orbits occur at relatively smaller radii, the stars on
these initially unperturbed orbits experience stronger distorting
forces than the progenitors of aligned orbits. (By ``progenitor'' of
an orbit, I imply that circular orbit which when subjected to the
perturbing potential, turns into the orbit at hand). This suggests
that the eccentricity of the anti-aligned orbits will in general be
higher than that of aligned orbits.  The more eccentric an orbit, the
greater is the radial excursion of the star.  Thus, we can expect that
the anti-aligned group in general will exhibit greater radial speeds
than the aligned group.  Initially circular orbits are distorted into
anti-aligned orbits in the region between CR and OLR by the perturbing
bar. Outside the OLR, circular orbits are deformed into aligned
orbits.  So, within the OLR, we expect many more anti-aligned orbits
than aligned orbits.  Outside the OLR, the converse is true.  On this
basis, we expect that the more populous clump inside the OLR
represents the anti-aligned family of orbits, while the more populous
clump outside the OLR represents the aligned family of orbits.

The velocity distributions change significantly with change in the
azimuthal location of the observer, even at the same radius.  Changes
in populations of the aligned and anti-aligned families can be
understood on the basis of orbital geometry. Figure~\ref{fig:samefam}
shows that the progenitors of anti-aligned orbits observed at a given
radius and at low azimuths are larger than the progenitors of the
anti-aligned orbits observed at the same radius but at higher
azimuths.  This size restriction on the progenitor at low azimuths
implies that its radius typically exceeds $\ROLR$ -- in which case,
the unperturbed orbit would be distorted into an aligned orbit rather
than an anti-aligned one. As $R$ gets progressively smaller, the
chance of the progenitor spilling outside the OLR reduces. Thus, at
radii close to the resonance, there is a dearth of anti-aligned orbits
observed at low azimuths. With increasing azimuth, the population in
the anti-aligned family grows.  As far as the aligned family is
concerned, it is again true that in the vicinity of the resonance
radius, at low azimuths there will be a scarcity of stars on aligned
orbits. This is due to the fact that the progenitors of aligned orbits
visible at low azimuths need to be smaller than those that distort
into aligned orbits seen at high azimuth. The size requirement
confines the initially circular orbit to $R<\ROLR$, in which case it
can only be distorted to an anti-aligned orbit. Again, the aligned
family becomes more populated as the observer moves away from the
resonance to higher radius. This is clearly shown in
Fig.~\ref{fig:standardmodel}, in which the clumps at an azimuth of
$65^{\circ}$ appear more populous than those at $25^{\circ}$.

\begin{figure*}
\centerline{
\psfig{figure=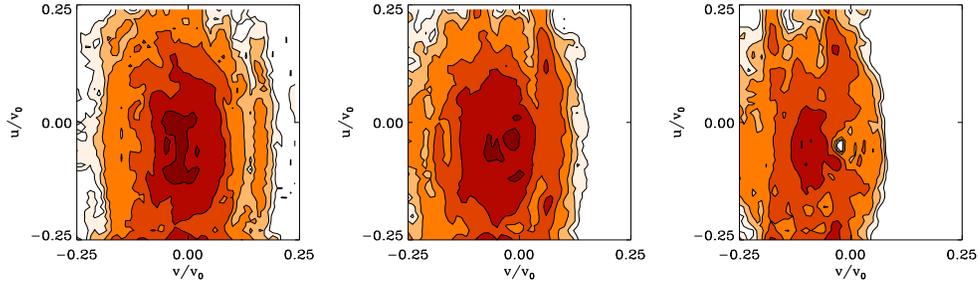,height=4.125truecm,width=13.2truecm}}
\caption{Plots of the velocity distributions in the ($U,V$) plane
after the growth of the standard spiral pattern in a warm exponential
disk. The chosen pattern speed of the spiral arm, places its ILR
inside the physical location of the OLR of an $m=2$ perturbation of unit
pattern speed. The left-most panel shows the distribution at this
physical location. The other panels represent the picture at
progressively increasing radii. All other parameters are the same as
in Figure~\ref{fig:exponentialmodel}.}
\label{fig:expouvsp}
\end{figure*}

Fig.~\ref{fig:expouvsp} represents the heliocentric velocity
distribution diagram obtained at the end of a simulation done with the
standard spiral pattern imposed on an equilibrium quasi-exponential
disk. The strength parameter $\zeta = 0.0155$.  The panels represent
the distributions at the different radial locations of the
observer. The left panel corresponds to the physical location of the
OLR of an $m=2$ disturbance (such as a bar). This location is outside
the ILR due to the spiral pattern used in the simulation, given a
ratio of $25:55$ between the pattern speeds of the two perturbers. The
middle and right panels are at progressively increasing radii outside
the resonance location. The azimuthal location of the observer is
$25^\circ$ for all the diagrams.

\begin{figure*}
\centerline{
\psfig{figure=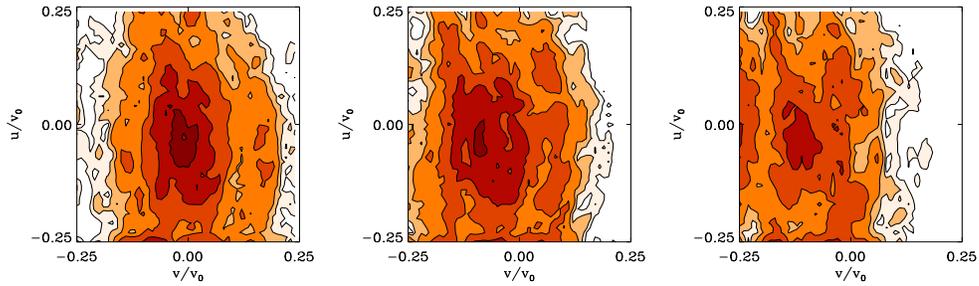,height=4.125truecm,width=13.2truecm}}
\caption{Plots of the velocity distributions in the ($U,V$) plane
after the simultaneous growth of a bar and the standard spiral pattern. 
Details are the same as in Figure~\ref{fig:expouvsp}. }
\label{fig:expouvspbar}
\end{figure*}

The velocity distribution diagrams obtained at the end of the
simulation done with the standard bar and spiral simultaneously
perturbing a quasi-exponential disk are represented in
Figure~\ref{fig:expouvspbar}. While the left panel represents the
distributions at the location of the OLR due to the bar, the other
panels are selected at progressively increasing radii. The panels
exhibit the distributions at azimuths of $25^\circ$.


\section{Velocity Dispersions}
\label{sec:disp}

\begin{figure*}
\centerline{
\psfig{figure=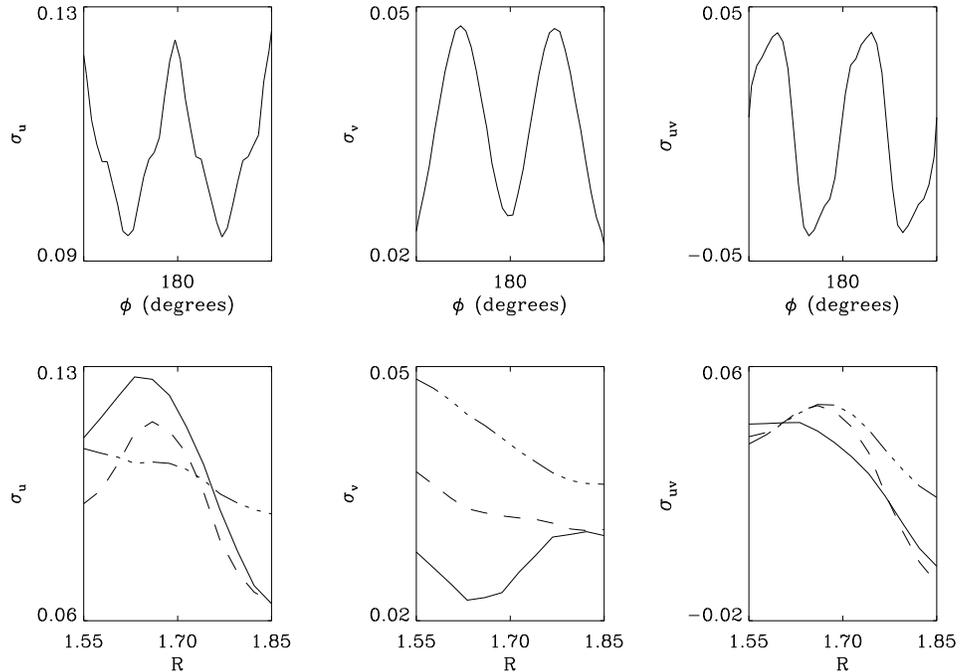,width=13.25truecm}}
\caption{Plots of the variation of the velocity dispersion with
position at the end of a simulation done with a bar imposed on the
standard Mestel disk. The upper panels show the variation of \sigmau,
\sigmav $\:$ and the cross-term \sigmauv, $\:$ with azimuthal angle,
at the resonance location. Here, the field due to the bar is about
$1.85\%$ that of the background disk.  The lower panels show the
variations with radius at three different azimuths, namely $25^\circ$
(solid line), $45^\circ$ (dashed line) and $65^\circ$ (dash-dotted
line).}
\label{fig:standarddisp}
\end{figure*}

\begin{figure*}
\centerline{
\psfig{figure=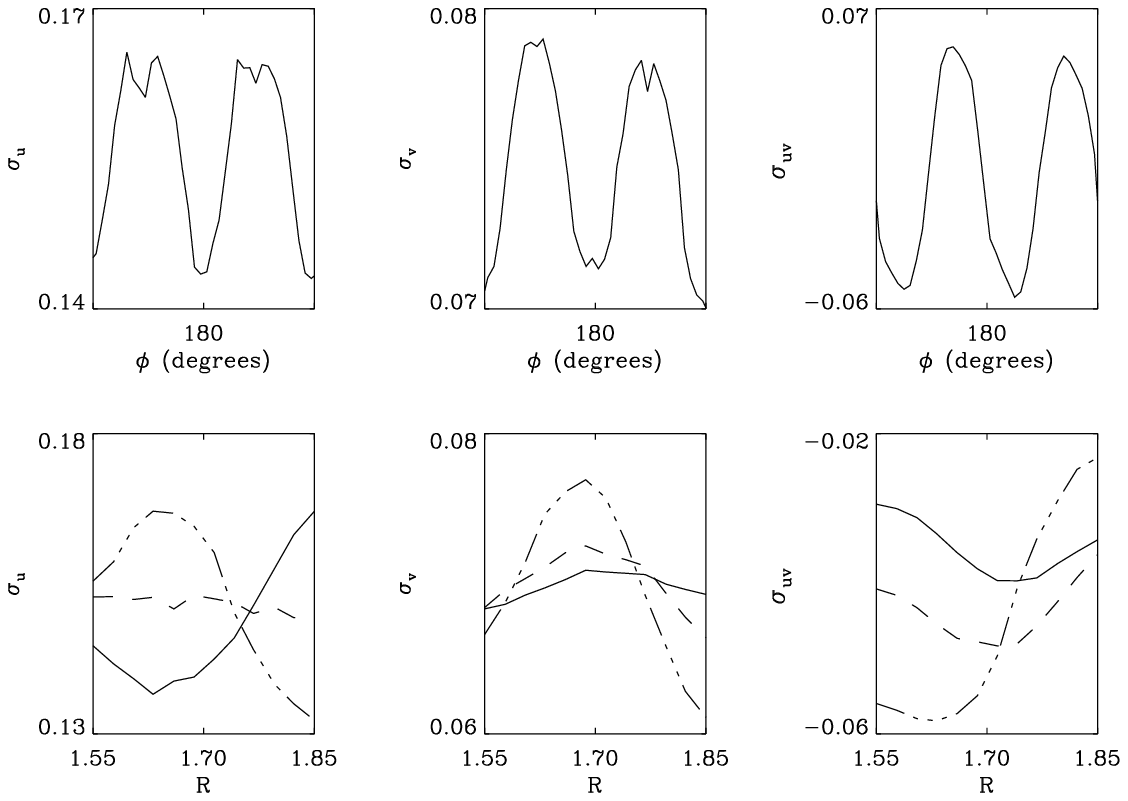,width=13.25truecm}}
\caption{Plots of the variation of the velocity dispersion with
position after the warm exponential disk has been perturbed by a bar. 
At the Outer Lindblad Resonance, the bar is about $2.4\%$ as strong
as the background disk. All details are as in Figure~\ref{fig:standarddisp}.}
\label{fig:expobardisp}
\end{figure*}

\noindent
The bisymmetric nature of the perturbation at the OLR location, is
clearly borne out in the Figures~\ref{fig:standarddisp} $\&$
~\ref{fig:expobardisp}. In fact, the $m=2$ symmetry comes out 
even more clearly when the bar perturbs a cold rather than a warm disk.
This is because the stars are more restless in the warmer disk.
The effect of the bar is most strongly imposed at the resonance
location.  Hence the radial velocity dispersion is expected to be
highest at this radius. As far as the dependence of \sigmau$\:$ on
azimuth is concerned, the perturbation due to the bar is strongest
along its major axis and weakest along its minor axis, at any
radius. Therefore we expect the radial velocity dispersion to be
highest at $\phi=0$ and lowest at $\phi=\frac{\pi}{2}$, at the
resonance. This is usually true as borne out by the panels showing the
variation of \sigmau$\:$with azimuth. It is reasonable to expect that
the total velocity dispersion, at any radius, will be given by the
depth of the potential at this location. Hence at any radius, the
gradient in \sigmau,$\:$ (with change in azimuth) is expected to have
the opposite sign to the change in \sigmav$\:$ as a function of
azimuth.  Thus tangential velocity dispersion is expected to be lowest
along the major axis and highest along the minor at $\ROLR$. However
these trends are challenged when the intrinsic stellar velocity
dispersion in the disk is high enough to impart sufficient energy to
some of the stars to disobey the perturbing potential.

\begin{figure*}
\centerline{
\psfig{figure=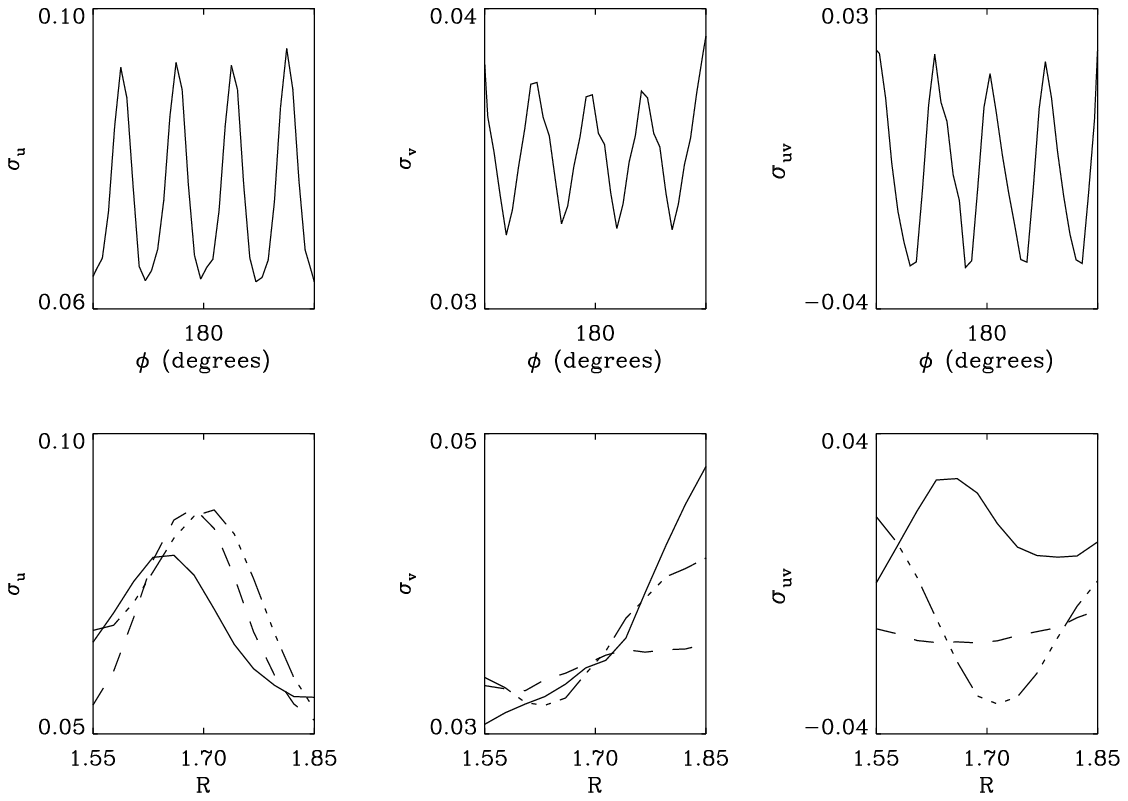,width=13.25truecm}}
\caption{As in Figure~\ref{fig:standarddisp} except that in this case,
the standard spiral pattern perturbed a cold Mestel disk.  The upper
panels show the variation of velocity dispersions with azimuthal
angle, at the location of the ILR due to the $4$ armed spiral pattern
(which corresponds roughly to the OLR location due to the bar since
the ratio of pattern speeds of the bar and the spiral pattern is 55:21
in this simulation). The gravitational field due to this spiral
pattern is about $1.5\%$ that of the background disk at this radius.}
\label{fig:standardspdisp}
\end{figure*}

The azimuthal dependence of the dispersions is shown at the ILR of to
the spiral pattern in Figure~\ref{fig:standardspdisp}. At this
radius, the potential of the logarithmic spiral achieves maxima at a
non-zero value of azimuth.  Therefore, radial velocity dispersion
peaks and tangential velocity dispersion dips at this azimuth rather
than at $0^\circ$ as in Figure~\ref{fig:standarddisp}. Again, as in
the bar simulations, these trends are not maintained in the warmer
disk.  As expected, the four-fold symmetry of the imposed spiral
pattern manifests itself in the variation of the velocity dispersions
with azimuth, at the ILR due to the spiral pattern (which is also the
OLR location due to the bar, approximately). The symmetry comes out
much more clearly in the simulation done with the colder Mestel disk
than the warm quasi-exponential disk. The symmetry is most marked at
the resonance location since the perturbation is felt most strongly
here.  As the spiral pattern is quite tightly wound, (i =
$15^{\circ}$) the perturbation appears to be approximately
axisymmetric.  Consequently, the effect of the spiral arms at
different azimuths and a given radius is almost the same.

%

\section{Heating Results}
\label{sec:locavgheat}

\noindent
The imposed perturbation heats the equilibrium disk depending on its
strength. One measure of the heating is the ratio of the final
location-averaged velocity dispersion to the average dispersion in the
equilibrium disk. These ``average'' dispersions are easily calculated
as follows. Once the orbits are recorded, the orbital data is gridded
according to location and the local velocity dispersion calculated.

\begin{figure}
\centerline{
\psfig{figure=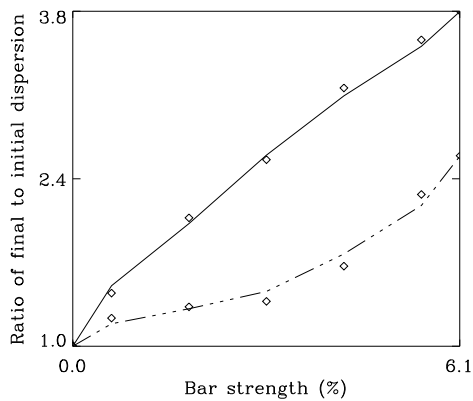,width=7.0truecm}}
\caption{Plot of the ratio of final to initial $\langle$\sigmau
$\rangle$ (unbroken line) and $\langle$\sigmav$\rangle$ (broken line)
in a cold Mestel disk, perturbed by a bar. The bar strength is the
ratio, (expressed as a percentage) of the gravitational fields due to
the bar and the background disk, at OLR. The unfilled circles represent
the calculated values of the change in the velocity dispersions for different
bar strengths, while the lines represent smooth fits to these data.
}
\label{fig:standardheat}
\end{figure}

\begin{figure*}
\centerline{
\psfig{figure=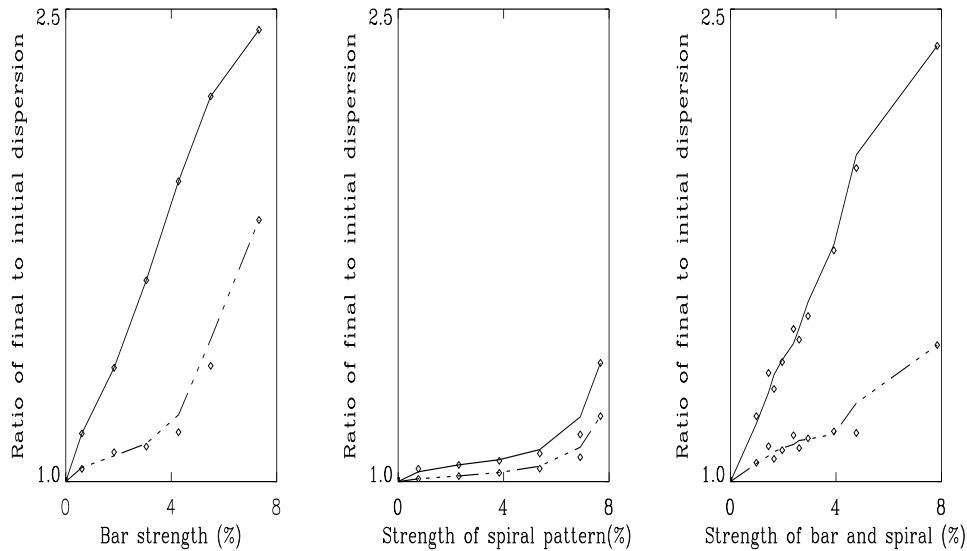,height=8.0truecm,width=13.25truecm}}
\caption{Plot of the ratio of final to initial $\langle$\sigmau
$\rangle$ (unbroken line) and $\langle$\sigmav $\rangle$ (broken line)
as a function of the ratio of the maximum field exerted by 
the perturbation to the field due to the disk ($\zeta$),
obtained at the end of simulations in which a warm quasi-exponential
disk is perturbed by a bar (left panel), the standard spiral pattern
(middle panel) and a bar and the standard spiral (right panel). The
ratio of the maximum radial force due to a quadrupole, a tightly wound
logarithmic spiral and the quadrupole $\&$ spiral together, (all of
unit strengths) to that of the background disk, at the OLR, is given
by $1:1.25:1.6$.}
\label{fig:expoheat}
\end{figure*}

Let the velocity dispersion at the $i^{\rm th}$ location bin be
$\sigma_{\rm i}$, the number of stars remaining in this location bin
after the growth of the perturbation be $n_{\rm i}$ and the arithmetic
mean of velocities of these $n_{\rm i}$ stars be $\mu_{\rm i}$. Let
there be $N$ such location bins in total. The average variance of all
the $\sum_{\rm i=1}^{N}{n_{\rm i}}$ stars is then given by:
\begin{equation}
\langle\sigma\rangle^2 = {\ds
                             {\frac
                              {\sum_{\rm i=1}^{N}
                               (\sigma_{\rm i}^2 + \mu_{\rm i}^2)n_{\rm i}}
                              {\sum_{\rm i=1}^{N}
                               {n_{\rm i}}}
                             }
                            } - 
                          {\left[
                                 {\ds
                                     \frac
                                     {\sum_{\rm i=1}^{N}
                                      {\mu_{\rm i}n_{\rm i}}
                                     }
                                     {\sum_{\rm i=1}^{N}
                                      {n_{\rm i}}
                                     }
                                  }\right]^2}
\end{equation}
The plots presented in Figure~\ref{fig:standardheat} show that the
effect of imposing the bar on the standard cold Mestel disk is to
elongate the radial velocity axis of the velocity ellipsoid more than
the tangential velocity axis. Thus, the ellipsoid is distorted due to
the perturbation. It is also observed in the figure, that heating
increases in general, as the bar grows stronger.
Figure~\ref{fig:expoheat} represents heating caused by the bar, the
spiral pattern and the two together, when they perturb a warm
exponential disk. Heating is plotted as a function of perturbation
strength $\zeta$.  The heating diagrams in Figure~\ref{fig:expoheat}
show that velocity dispersions generally increase with increasing
perturbation strength. Tangential heating is always less than radial
heating at all perturbation strengths.  It can be seen in
Figure~\ref{fig:expoheat} that for a given ratio between the
gravitational field due to the perturber and the disk, in general
heating is maximum for a given perturbation strength, when the bar
acts alone.

The spiral pattern alone is the least efficient heating mechanism. In
fact, around the ILR due to the spiral arms, the location-averaged
disk heating is less than that around the OLR due to a bar of the same
perturbation strength $\zeta$. A spiral pattern that is
quasi-stationary cannot efficiently heat the stellar disk. This
follows from the smooth, largely unchanged nature of the velocity
distribution away from the resonances.

\section{Surface Density Profile}

\noindent
I have investigated the response of the disk to the perturbation by
plotting the number of stars at a given azimuthal location as a
function of radius. This number is referred to as the ``occupation
number'' in the paper.  In fact, the integration of the surface density
over a small local patch of the disk, provides the occupation number
at that location. The occupation number in an equilibrium Mestel disk
is a constant, independent of radius, while it has a more complicated
dependence on radius for a disk with an initially exponential surface
density profile.

\subsection{Bar}
\begin{figure}
\centerline{
\psfig{figure=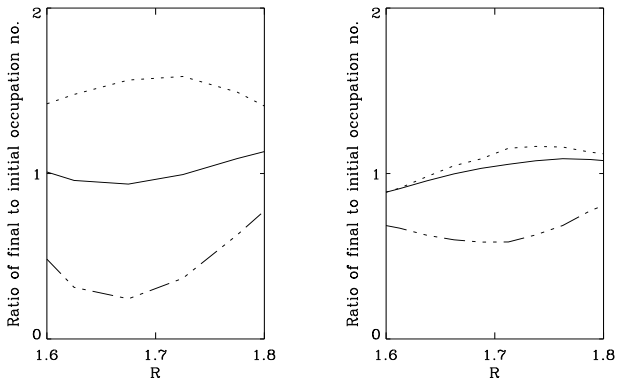,height=5.5truecm,width=9.0truecm}}
\caption{Plot of the variation of the final to initial occupation
numbers with radius when a bar is used to perturb the equilibrium
disk; the panel on the right corresponds to the standard
quasi-exponential disk and that on the left to the standard Mestel
disk.  The solid line represents the azimuthally averaged ratio of the
final and initial occupation numbers while the dotted and the
dash-dotted lines represent the ratio at azimuths of $90^\circ$ and
$0^\circ$ respectively. In this case, $\zeta = 1.8 \%$.}
\label{fig:barsurface}
\end{figure}

\noindent
The ratio of the final to the initial occupation numbers obtained at
the end of simulations done with a bar, imposed on the standard Mestel
and the quasi-exponential disks, is plotted as a function of the
radial location of the observer, at different azimuths in
Fig.~\ref{fig:barsurface}.

%

It is observed that there is a depletion of stars from the region
along the major axis of the bar ($0^{\circ}$) and an accumulation
along the minor axis ($90^{\circ}$); this depletion and accumulation
is strongest near the resonance radius. The variation in the
occupation number is basically similar for both the exponential disk
as well as the Mestel disk.  As has been explained in
Section.~\ref{sec:baruv}, there is a dearth of stars on both aligned
and anti-aligned orbits at low azimuths, near the resonance. This
shows up in the occupation number plot as a depletion at
$\phi=0^\circ$, near the resonance location. Near the resonance, at
high azimuths, both the orbital families are strong; however the
aligned family gets thinner as radius decreases and the anti-aligned
is progressively debilitated as radius increases. Thus there is a peak
in the occupation number along the minor axis of the bar, near
$\ROLR$.

When the disk is much hotter, more stars can travel from outside the
resonance to inside and vice versa. Thus the cause for the depletion 
and accumulation of stars along certain directions is less prominent
in a warmer disk. Also, when the initial occupation number  
profile of the background disk is more exponential than flat, 
many more stars can come over from radii $<\;\ROLR$, thus keeping
the anti-aligned family well populated even at the low azimuths. 
These factors imply the flatter nature of the occupation number profiles
when the bar is imposed on a warm exponential disk.
\begin{figure}
\centerline{
\psfig{figure=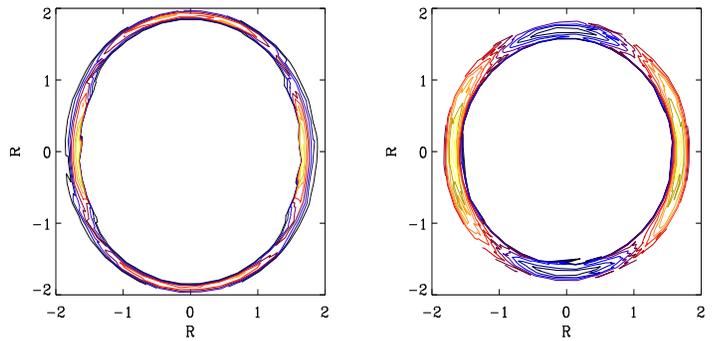,height=5.0truecm,width=10.0truecm}}
\caption{Isodensity contours in the vicinity of the OLR location when
the bar perturbs the cold disk (left panel) and the warm disk (right
panel). All other parameters are as in
Figure~\ref{fig:barsurface}. The bar is along the horizontal.}
\label{fig:boxy}
\end{figure}

The ability of the bar to distribute stellar matter at the resonance
location, across azimuths is also manifest in the isodensity contours
turning boxy. Figure~\ref{fig:boxy} depicts isodensity contours for
which the occupation number calculations were shown in
Figure~\ref{fig:barsurface}. As is evident from the plots in this
latter figure, the dip in the occupation number at low azimuths occurs
just inside the OLR radius in the cold disk and just outside this
location in the warmer disk. This explains why the contours shown in
Figure~\ref{fig:boxy} are anti-aligned to the bar in the colder disk
(left panel of Figure~\ref{fig:boxy}) and aligned to the bar major
axis in the warmer disk (right panel). In both cases, the contours are
boxy though the deviation from an elliptical shape is more conspicuous
in the cold disk case. The effect of the bar is less strong in the
warm disk case as is evident from Figure~\ref{fig:barsurface}, thus
contributing to the boxiness of the isodensity contours to be less
marked.

\subsection{Spiral Pattern}
\label{sec:spocc}
\begin{figure}
\centerline{
\psfig{figure=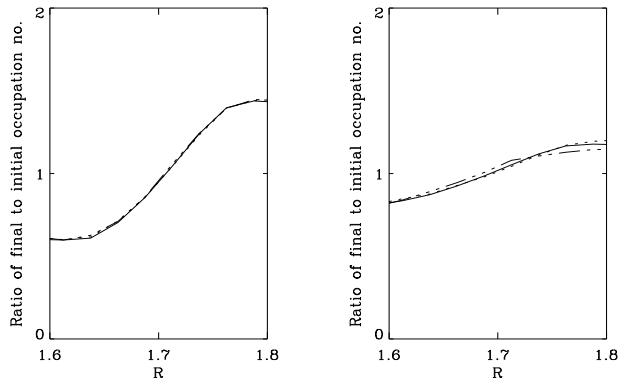,height=5.5truecm,width=9.0truecm}}
\caption{Plot of the variation of the final to initial occupation
numbers with radius when the equilibrium disk is perturbed by a
spiral pattern. The choice of the pattern speed of this spiral,
implies that the ILR of this spiral pattern lies almost on top of the
OLR due to an $m=2$ disturbance, such as a bar. In this case, $\zeta
= 1.125\%$. All other details are as in Fig.~\ref{fig:barsurface}. }
\label{fig:spsurface}
\end{figure}

\noindent
Fig.~\ref{fig:spsurface} shows the variation in the ratio of the final
occupation number to the same in the initially unperturbed equilibrium
disk, at selected azimuths, with radius.  As was explained in
Section~\ref{sec:disp}, the standard pattern is approximately
axisymmetric. This implies that the effects of the spiral arms will be
nearly isotropic on the disk. This explains why occupation number is
almost the same at all azimuths. It appears from
Fig.~\ref{fig:spsurface} that the spiral pattern tends to deplete
stars away from lower radii and allow them to accumulate at higher
radii. This can be attributed to more efficient angular momentum
transfer from lower to higher radii, by the spiral as compared to the bar.
In fact, the ILR due to the spiral, is an ``emitter'' of angular momentum, 
(\cite{donaldagris}). Thus we expect to find a re-distribution of disk
surface density towards higher radii, in the immediate neighbourhood
of the ILR. The spiral pattern is not efficient in such re-distribution of
matter at radii other than the resonances. 
In the warm quasi-exponential disk, stars have higher random motion
and the effect of the perturbation is manifest to a lesser
degree. Thus occupation number varies more with azimuth in the right
panel in Fig.~\ref{fig:spsurface} than in the left.

\subsection{Bar and Weak Spiral Pattern}
\begin{figure}
\centerline{
\psfig{figure=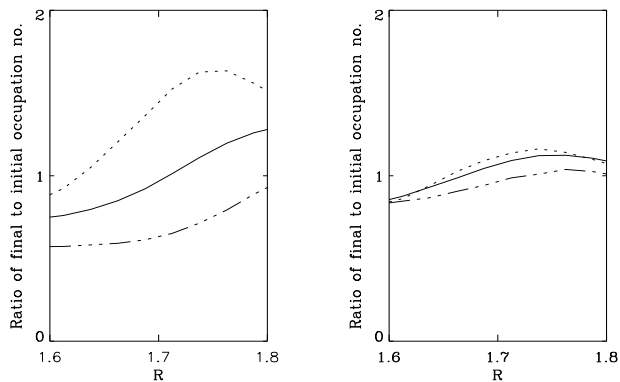,height=5.5truecm,width=9.0truecm}}
\caption{Plot of the variation of the final to initial occupation
numbers with radius when the equilibrium disk is perturbed by the bar
and the standard spiral pattern, jointly. In this case, $\zeta =
3.8\%$. All other details are as in Fig.~\ref{fig:spsurface}. }
\label{fig:spbarsurface}
\end{figure}
\noindent
In Fig.~\ref{fig:spbarsurface} the ratio of the final occupation
number to the initial occupation number in the unperturbed equilibrium
disk, is plotted as a function of radius at chosen azimuths. Here the
disk is perturbed by the bar and the spiral pattern jointly.  We have
seen in Fig.~\ref{fig:barsurface} that there tends to be a depletion
of stars from the region along the major axis of the bar and an
accumulation along the minor axis. The spiral pattern appears to drag
stars away from lower radii, outwards. Both these effects are
superimposed to produce the picture presented in
Fig.~\ref{fig:spbarsurface}.

\section{Growth and Dissolution of a Bar and a Spiral Pattern}

\noindent
It is interesting to examine the extent to which the disk is able to
retain memory of the phase space features caused by non-axisymmetric
perturbations.  This spurred the study of the effects of an imposed bar
and the standard 4-armed spiral pattern, on disk stellar kinematics,
as a function of the time periods over which the perturbation is
grown ($T_1$) and dissolved ($T_3$) as well as the time over which the
perturbation strength remained saturated to its maximum value
($T_2$). In these simulations, the ``growth time'' is equated to the
``dissolution time'' of the perturbation, ($T_1=T_3$). $T_2$ is
referred to as the ``steadiness time''. I first present the effects
of a bar on a cold Mestel disk and follow this up with an
investigation of the effects brought about by a bar and spiral pattern
on a warm exponential disk.

\subsection{Mestel Disk}
\label{sec:mesdissbar}

\begin{figure*}
\centerline{
\psfig{figure=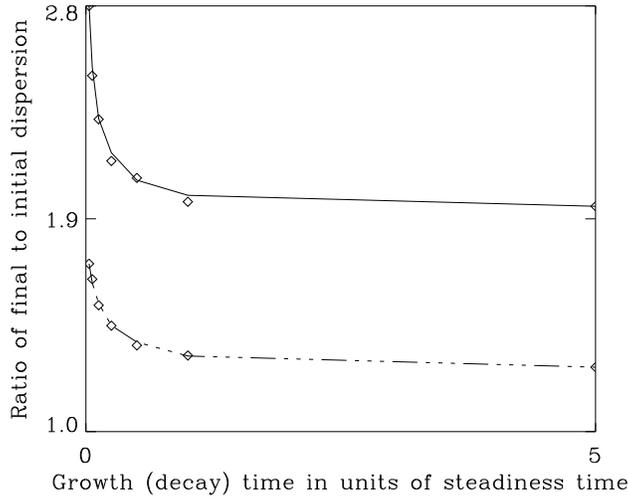,width=9.0truecm}}
\caption{Plot of variation in the disk heating caused by a bar
which is grown and subsequently dissolved in the standard Mestel disk.
The ratio of the final to the initial average radial velocity
dispersion (solid line) and the same for the tangential velocity
dispersion (broken line) is plotted as a function of the ratio of the
``growth time'' (or ``dissolution time'') to the ``steadiness time'',
($T_1/T_2$). The heating, as calculated at a value of the ratio
$T_1/T_2$, is marked by unfilled circles, while the lines are smooth
fits to these data. The bar strength is allowed to reach a maximum
value that corresponded to $1.2\%$ of the background field at the OLR.
}
\label{fig:messdiss}
\end{figure*}

Fig.~\ref{fig:messdiss} displays the location-averaged radial and
tangential heating produced by a bar that is imposed on a cold Mestel
disk and subsequently dissolved. It is observed that the
bar-induced heating initially decreases rapidly with increasing growth
time for a constant $T_2$, but very soon the rate of decrease becomes
nearly zero. The degree of distortion of the velocity ellipsoid
remained a constant approximately.

The trends manifest in Figure~\ref{fig:messdiss} can be understood in
the light of the following discussion. When the bar is being grown and
dissolved quickly, it is expected that stellar velocities in the disk
will be rendered high. The slower the growth (and dissolution), the
more nearly adiabatic and reversible are the conditions; so the
smaller the final heating. Thus, there is higher disk heating at
smaller values of $T_1$.

As the bar is grown (and killed) progressively slowly, there appears to be a
range of the growth (and dissolution) time over which heating remains
nearly a constant; (heating drops with even slower growths). This
range in $T_1$ has been captured in Fig.~\ref{fig:messdiss}. As the
bar is dissolved more slowly, phase mixing becomes more efficient
in the disk; thus it is not surprising that heating drops with
increase in $T_1$ and $T_3$. Phase mixing is more efficient in the
warm quasi-exponential disk than in the initially cold Mestel
disk. Therefore we would expect a faster drop in disk heating with
$T_1$ (=$T_3$) in the former than the latter configuration. This is borne
by a comparison of Fig.~\ref{fig:messdiss} and Fig.~\ref{fig:expodiss}.

I have noted in the experiments that the time over which the bar is
allowed to remain at its maximum strength ($T_2$) does not affect
heating, for fixed $T_1$ and $T_3$, as long as $T_2$ is sufficiently
high. The disk heating recorded after the dissolution of the bar is
determined by the efficiency of phase mixing. If $T_2$ is higher than
a few bar rotation periods, the orbital eccentricities induced by the
bar saturate and any further increase in $T_2$ will not distort the
orbits. Roughly speaking, orbital eccentricities can be used
as indicators of disk heating. Thus, enhancing the ``steadiness time''
beyond a point will not bring about any change.

As expected, heating depends sensitively on the bar strength for given values
of $T_1$ (=$T_3$) and $T_2$.

\subsection{Warm Quasi-exponential Disk} 
\label{sec:expodiss} 
\begin{figure*}
\centerline{
\psfig{figure=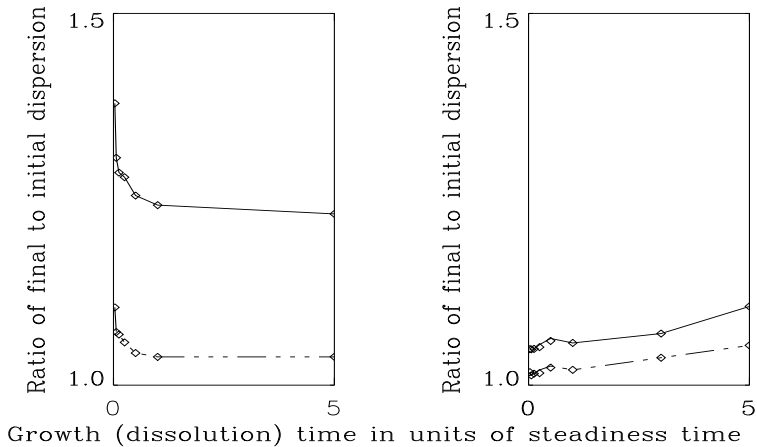,height=6.5truecm,width=11.0truecm}}
\caption{Plot of variation in the disk heating caused by a bar (left
panel) and the standard spiral pattern (right panel), which are grown
and subsequently dissolved in the warm quasi-exponential disk. The
ratio of the final to the initial average radial velocity dispersion
(solid line) and the same for the tangential velocity dispersion
(broken line) is plotted as a function of the ratio of the growth (or
dissolution) time to the growth time, ($T_1/T_2$). The lines are
actually smooth fits to the heating data, which is represented by
unfilled circles. The bar contributes a maximum field that is $1.8\%$
of the background field, while the maximum field due to the spiral
pattern is only $0.75\%$ of the field due to the Mestel potential, at
the OLR due to the bar.}
\label{fig:expodiss}
\end{figure*}
\begin{figure*}
\centerline{
\psfig{figure=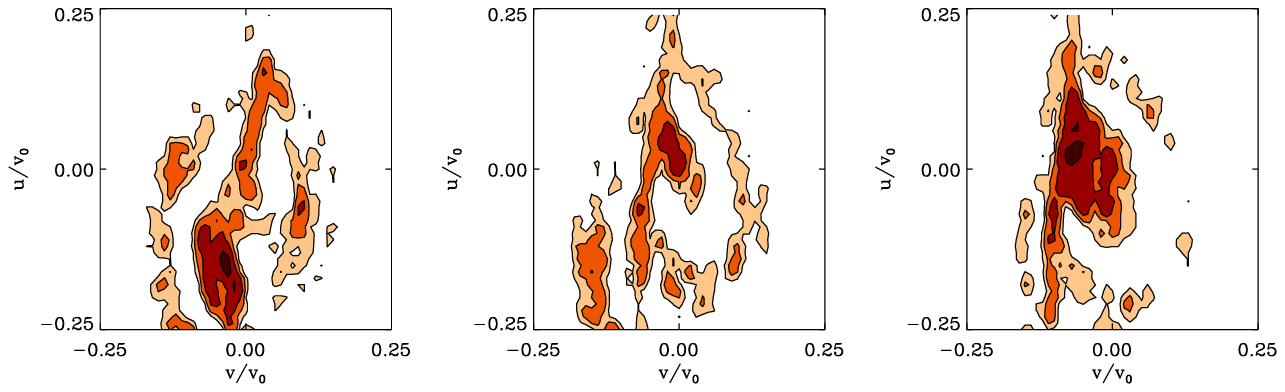,height=4.125truecm,width=13.25truecm}}
\caption{Plot of velocity distribution diagrams at the end
of simulations in which a bar is grown and subsequently dissolved in
the warm exponential disk. The panels represent observer radial
locations which are inside, at and outside $\ROLR$, at an azimuth of
$25^{\circ}$. The velocity distribution is maximum at the centre and
decreases outwards in these plots, as indicated by the gradient in the
shading. The bar used in this simulation implies a maximum field that
is $1.8\%$ of that of the background disk, at OLR. The growth (dissolution)
time is 5 times the steadiness time, in this simulation.}
\label{fig:uvdissexpbar}
\end{figure*}

Fig.~\ref{fig:expodiss} explores the disk heating that is caused when
a bar (left panel) or a spiral pattern (right panel) is grown in a
warm exponential disk and is then dissolved. The location-averaged
radial and transverse heatings are plotted as functions of the ratio
$T_1/T_2$. The degree of heating due to the spiral is much lower than
that due to the bar in general.  

Simulations were done in which a bar was grown and dissolved in a warm
quasi-exponential disk. Heliocentric velocity distributions were
recorded at the end of such simulations at different locations of the
observer. In Fig.~\ref{fig:uvdissexpbar}, a set of velocity diagrams
representing the structures in velocity space after dissolution of the
bar at the end of moderately long growth and dissolution times, are
shown in the ($U-V$) plane at locations inside, at, and outside the
OLR due to the bar at an azimuth of $25^{\circ}$. These structures
obtained with an initially warm quasi-exponential disk are reminiscent
of the features observed after a bar has been grown in a cold disk,
(shown in Fig.~\ref{fig:standardmodel}), insofar as the two stellar
families of oppositely oriented orbits are concerned. However,
velocity distribution diagrams obtained at the end of simulations done
with smaller values of the ratio ($T_1/T_2$), exhibit structure that
is less well-defined.

The visual appearance of the structures in Fig.~\ref{fig:uvdissexpbar}
suggests that they resemble the velocity
distribution plots, obtained after stirring a cold disk with a bar,
except that now, there are regions of phase space which are not
occupied. Such ``holes'' are observed in the distributions
irrespective of how slowly the bar is grown and dissolved.  This
observation is due to the fact that initially circular orbits that lay
very close to the OLR, can become chaotic, by the introduction of the
bar. In an action diagram, in the immediate vicinity of the resonance
location, such orbits are known to occupy discontinuous patches,
(\cite{binney84}). The missing actions correspond to missing phase-space
volumes. Thus, parts of the recorded velocity distributions are noted
to be empty. These ``holes'' will be observed even when the perturbation
potential is changed in the adiabatic limit.
\begin{figure*}
\centerline{
\psfig{figure=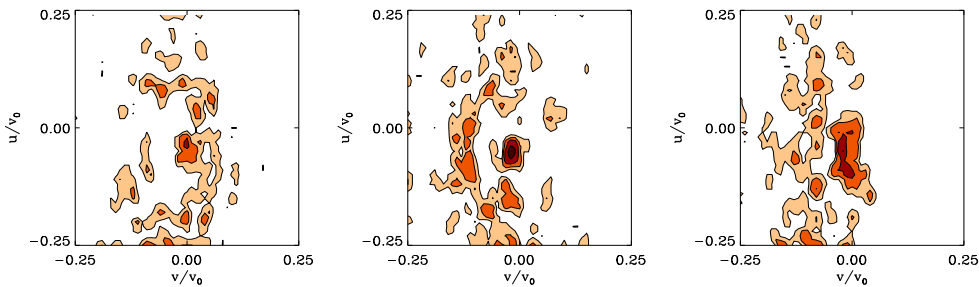,height=4.125truecm,width=13.25truecm}}
\caption{Plot of velocity distribution diagrams at the end
of simulations in which the standard 4-armed spiral pattern
is grown and subsequently dissolved in the warm exponential disk.
The panels represent observer radial locations which are inside,
at and outside the ILR due to the spiral pattern, at an azimuth of 
$25^{\circ}$. The maximum field contributed by the spiral pattern is
about $1.5\%$ of that due to the background disk, at the ILR. In this
simulation, $T_1/T_2=8$. }
\label{fig:uvdissexpsp}
\end{figure*}

The velocity distribution diagrams in the vicinity of the ILR due to
the standard spiral pattern (which corresponds very closely to the
location of the OLR due to the bar), as recorded by an observer at the
Sun, at the end of a simulation in which the spiral pattern is grown
and then dissolved in a warm exponential disk, is shown in
Fig.~\ref{fig:uvdissexpsp}. Structures in the velocity space can be
recognised from these figures. Again, as for the case of the bar,
these distributions resemble velocity distribution plots, obtained by
stirring a cold disk with a spiral pattern, except that the current
plots exhibit missing patches of phase-space volume.

\section{Conclusions}

\noindent
This paper presents test particle calculations showing the effects of
bars and spiral patterns on the outer parts of galactic disks.  The
effects of the perturbations on the properties, such as the velocity
dispersions and the stellar density, have been examined in detail.
The changes are most pronounced at the resonance locations of the bars
and spirals.

The simulations begin from idealised infinitesimally thin disks with
smooth velocity distributions that are generated from the
collisionless Boltzmann equation. Even so, the effects of the bar and
spiral distortions cause the final velocity distributions to be far
from smooth. Tightly-wound, many-armed spiral patterns do least damage
to the smoothness of phase space, whereas bars cause much more
disturbance.

The effects of the imposed perturbations are enumerated as follows.
\begin{enumerate}
\item
In the model in which a cold disk is perturbed by a bar, Kalnajs'
mechanism causes definite bimodality at locations near the OLR. This
effect is very clean in the cold disk, and the clumps can be readily
identified with the families of aligned and anti-aligned orbits. However
if a warmer disk is so perturbed, then the structure shows complex and
multiple peaks, caused by the scattering of the original orbits onto
resonant families. Spiral patterns {\it do not} cause bimodality.

\item The resonances can be locations of anomalous heating. This
effect is most evident in results obtained from simulations done with
completely cold disks, in which the imposition of a bar or a spiral
pattern causes a peak in the radial component of the velocity
dispersion \sigmau $\:$ at the resonance. When the disk is hotter,
then the velocity dispersions are increased by any perturbation, but
\sigmau $\:$ may either increase outward or decrease outward depending
on the azimuth.

\item The resonances may also be locations where the disk isophotes
are distorted from ellipses and are boxy. The effect of a bar is found
to be to cause stars to become depleted from the major axis and
accumulate on the minor axis at the OLR. Boxiness of isophotes has
been associated with the presence of a bar by other workers, both in
numerical (\cite{patsis03}) and observational studies (\cite{pohlen03}),
but only till about the bar length. The occupation number
investigation reveals that the bar can act in a similar way, even in
the outer disk, around the OLR. Spirality causes stars at the ILR (of
the spiral) to gain angular momentum and drift outward. When a
spiral pattern and a bar are made to act in conjunction, the degree of
the inflicted boxiness on the isophotes is less than when the bar is
acting alone.

\item Computations have been carried out to calculate the typical
heating effects of bars and spiral patterns in the outer parts of
galactic disks. For a typical galaxy like the Milky Way, we expect the
growth of the bar to have roughly doubled or quadrupled the velocity
dispersion, depending on whether the initial configuration was hot or
cold. For disks presently like the Milky Way (with \sigmau $\sim
45$\kms), the radial velocity dispersion is increased by a factor
$f\approx 1 + 0.19 \zeta_{\rm bar}$. Here, $\zeta_{\rm bar}$ is the
ratio of the maximum gravitational field of the bar to that of the
disk at the OLR. Spiral features are found to be much less efficient
at heating the outer parts of galactic disks.

\item Runs in which bars are grown and dissolved indicate that the
heating that accompanies the growth cannot be removed immediately on
dissolution. A very rapid growth and dissolution of a bar the size of
the Milky Way's in an initially cold disk may cause the velocity
dispersion to more than quadrupole in value. A less massive bar has a
smaller effect, as does a hotter disk and longer growth (and
dissolution) periods.

\item Another legacy left behind by the dissolved perturbers is on the
velocity distributions. When the perturbers are grown and subsequently
dissolved in a warm disk, the velocity distributions resemble those
obtained after the perturbation reaches maximum strength in an
initially cold disk, except that ``gaps'' appeared in the velocity
space. These are caused by missing orbits that cannot retrace their
steps once the perturbation is removed. This phenomenon has been
earlier noted by Binney \& Spergel (1984).
\end{enumerate}

The simulations reported in this work attempt to understand the effect
that imposed perturbations have on galactic disks, in general. A
future paper is planned in which the models studied herein are
explored in depth with the aim of replicating the outer parts of the
Milky Way disk. In particular, the velocity distribution in the Solar
neighbourhood is sought and compared to that observed by Hipparcos;
relevant Milky Way parameters are extracted from such an exercise.

A point that merits discussion is the extent to which chaos is present
in the phase space. In particular, it will be interesting to identify
regions of the phase space where chaos dominates and to associate the
fraction of chaotic orbits with the strength of the imposed
perturbation. This is fully achieved only via a rigorous treatment of
the orbits for the presence of chaos. This has not been taken up in
this paper; however, in the future paper mentioned in the last
paragraph, a surface of section analysis of orbits from specific patches
of the phase space is done. (The orbits being two dimensional, such a
simple analysis is deemed sufficient). 

It is indeed important to know if the phenomena studied in this paper
are caused by the action of the imposed non-axisymmetric perturbations
or chaos. \scite{contopoulos89} suggest that the region between
corotation and the outer 4/1 resonance is stochastic, when the imposed
perturbation is a bar or a spiral. However, the radial band of
interest in this paper is the immediate vicinity of the OLR of an
$m=2$ perturbation that has been assigned a pattern speed of
unity. Thus, in our scale free units, this region typically extends in
radius from 1.55 to 1.85. The -4/1 resonance is however located at
about $R=1.3$ which places it too far inside to overlap with the region
of interest in this paper. Hence inside this region of investigation,
there is no preset reason to apprehend the presence of chaos, except
perhaps the locations that are very close to the -2/1 resonance.
\footnote{Another interesting paper in this context is \scite{contopoulos88}
which presents an analysis of orbits in a barred galaxy, but within
the inner 4/1 resonance.}

It is possible that perturbers of different strengths, introduce
different degrees of chaos in the disks with different coldness
parameters. It is likely that stronger the perturbation, orbits are
more susceptible to becoming chaotic. Also, we have consistently
noticed that the effects of a perturber on a cold disk is similar in
nature to that on a hot disk with the exception being in terms of the
strength of the effect; the effect of the perturber is relatively more
prominent in the colder disk. For example, the heating caused by a bar
is nearly double in the colder disk than in the warmer disk. Also a
bar sculpts out the region along its major axis nearly thrice more
efficiently in a colder disk than in a warmer disk, in the vicinity of
the OLR.  These results appear to indicate that, at least in the
vicinity of the resonance locations, the perturber induces more orbits
to become chaotic in the hotter disk. When the background disk is
colder, the orbits are less susceptible to turn chaotic and the effect
of the perturber is more strongly felt. A figurative analogy can be
drawn with the case of throwing a pebble into a placid lake which
causes distinct ripples, while there is almost no effect visible if
the pebble is thrown into a rough sea.

In the section that dealt with the heating results,
(Section~\ref{sec:locavgheat}), we noted the low heating efficiency of
the spiral pattern.  A spiral pattern is very efficient in
distributing angular momentum through the disk, across a radial
range. However, a tightly wound spiral, like the one used in the
models, is almost axisymmetric effect and cannot do much in
distributing stars across azimuths, at any radius. This is unlike the
behaviour of a bar, which can move stars from along its major axis
towards its minor axis, most efficiently at the resonance locations.
The efficacy of the bar to distribute stars likewise decreases as one
moves away from the resonance. Hence, in the immediate vicinity of the
resonance location, there is significant heating produced by the
bar. In lieu of such a dynamical mechanism being in operation in the
case of the spiral, the level of heating induced by a spiral pattern
is low compared to a bar.

The ``occupation number'' results show the effect of the perturbers to
distribute stellar material across the disk. The advantage of
displaying this effect in terms of the ``occupation number'' plots,
over plots of isodensity contours is that the former type of display
offers direct access to quantified estimates of the effect in the
different cases. It has been noted in a number of N-body studies of
barred galaxies that isophotes turn boxy in the presence of the bar
(\cite{patsis03}), but most of these studies are carried on till radii
that is about the bar length. The occupation number calculations
presented above show that the bar can similarly bolster regions along
its minor axis at the expense of regions along its major axis, even in
the outer disk (at the OLR). There is a recent observational report of
peanut shaped outer isophotes in the galaxy UGC 7321
(\cite{pohlen03}). The boxiness of the outer isophotes are explained
in terms of a large bar; the bar is construed to be large in the light
of the reports currently available in the literature that connect
boxiness with the presence of a bar, at locations around the bar end.
However, the current work indicates that such a phenomenon is also
possible further out, at the OLR of the bar. From the radial location
of these observed boxy isophotes, the bar parameters could potentially
be modelled.

This work has shown the dangers of assuming smooth distribution
functions as advocated by Jeans theorem. In the outer parts of
galactic disks, there are many processes that can cause substructure
in velocity space. The next generation of an astrometric satellites
like FAME and GAIA have the capabilities to map outer velocity space
with unprecedented precision.  Many of the phenomena that are
described in the paper will come unto the purview of observations
within the next decade.  We are already in possession of full
kinematic information on a substantial number (14,139) of the nearby
stars from the latest report of the Geneva-Copenhagen survey of the
Solar neighbourhood, (\cite{nordstrom04}). Analysis of this treasure
trove of data promises to exemplify the phenomenology discussed above.

\section*{Acknowledgments}
{I would like to thank my Ph.D. supervisor, Dr. Wyn Evans who was my
collaborator in this work. His suggestions and comments made this
paper possible. I also thank my other Ph.D. supervisor, Dr. Prasenjit
Saha for his suggestions that helped me respond to the
referee's queries. I would like to acknowledge the Felix
Scholarship that paid towards my maintainence and tution in Oxford
where this work was initiated.}

\newpage
\bibliographystyle{mnras}
\bibliography{dc.bib}

\end{document}